# Orbital Hall effect assisted field-free perpendicular magnetization switching

Zelalem Abebe Bekele[1,2,†], Yuan-Yuan Jiang[3,4,†], Kun Lei[1,2], Xiukai Lan[1], Xiangyu Liu[1,2], Hui Wen[1,2], Ding-Fu Shao[3,*], and Kaiyou Wang[1,2,*]

[1] *State Key Laboratory of Semiconductor Physics and Chip Technologies, Institute of Semiconductors, Chinese Academy of Sciences, Beijing 100083, China*

[2] *Center of Materials Science and Optoelectronics Engineering, University of Chinese Academy of Sciences, Beijing 100049, China*

[3] *Key Laboratory of Materials Physics, Institute of Solid State Physics, HFIPS, Chinese Academy of Sciences, Hefei 230031, China*

[4] *University of Science and Technology of China, Hefei 230026, China*

* Corresponding Email: dfshao@issp.ac.cn, kywang@semi.ac.cn

†: These authors contributed equally to this work

## ABSTRACT

Spin-orbit torques (SOTs) generated through the conventional spin Hall effect (SHE) and/or Rashba-Edelstein effect offer potential for magnetization manipulation. However, deterministic switching of perpendicular ferromagnets via SOTs requires a strong symmetry-breaking perturbation, typically an external magnetic field. Here, we demonstrate that field-free SOT switching of perpendicular magnetization can be facilitated with the assistance of the orbital Hall effect (OHE). Using a representative Co/PtGd bilayer SOT device, we find that while the planar Hall effect (PHE) generates a finite out-of-plane damping-like torque, representing a lateral symmetry breaking, the SHE-induced torque achievable at practical current density is insufficient to switch the perpendicular magnetization. Incorporating a Mo underlayer and exploiting its strong OHE can amplify the in-plane damping-like torque via orbital-to-spin conversion, enabling efficient field-free deterministic switching without complex device geometries or low symmetric spin sources, providing a straightforward and scalable strategy for achieving high-speed and low-power spintronics.

## I. INTRODUCTION

Spintronics employ spontaneous magnetic states in materials for encoding information in nonvolatile memories, logic devices, and neuromorphic computing.[1,2] A high-performance spintronic device requires the efficient manipulation and detection of magnetic states, which is crucial for write-in and read-out operations. The Ferromagnets with perpendicular magnetizations are desirable for high-density applications. Thus far, efficient electrical manipulations of perpendicular ferromagnets remain a challenging problem. A promising approach is the spin-orbit torque (SOT), where an in-plane charge current $\boldsymbol{J}$ ($x$ direction) generates a nonequilibrium spin polarization $\boldsymbol{p}$, which exerts a damping-like torque $\boldsymbol{\tau}\sim\boldsymbol{m}\times(\boldsymbol{m}\times\boldsymbol{p})$ on the magnetization $\boldsymbol{m}$ of a ferromagnetic layer (FM) in a SOT device.[3,4,5] In the conventional spin Hall effect (SHE)[3,4] or the Rashba-Edelstein effect (REE)[5,6] that is widely used in the SOT devices, the generated $\boldsymbol{p}$ is usually aligned to the direction of $\boldsymbol{J}\times\boldsymbol{z}$, i.e. the $p_y$, due to the symmetry constraint, where $\boldsymbol{z}$ is a direction normal to the film plane. However, the switching by the associated in-plane damping-like torque $\tau_y\sim\boldsymbol{m}\times(\boldsymbol{m}\times p_y)$ is inefficient and non-deterministic. First, a large current is required to generate a sufficiently strong $p_y$ and the associated $\tau_y$, since $\tau_y$ need to overcome the precession induced by the anisotropy field.

Second, the $\tau_y$ can only pull $\boldsymbol{m}$ toward $y$ direction while applying the current, and the relaxation of $\boldsymbol{m}$ cannot be controlled when the current is released.[7] This is because the states of $\boldsymbol{m}$ pointing along +$z$ and -$z$ directions are degenerate in a perpendicularly magnetized FM film. Although an in-plane electric field breaks most symmetry operations connecting the two states, a mirror symmetry $M$ parallel to the electric field and perpendicular to the film plane may be preserved. Therefore, an additional perturbation to break $M$ is thus required to realize a deterministic switching. There are typically two paradigms for this sake. The first one utilizes an in-plane assisted magnetic field, which can be either an external field[8,9] or an exchange bias field from an adjacent antiferromagnetic film.[10,11] The second one introduces the lateral asymmetry of the FM film, such as using the FM film with a wedged shape,[12] a lateral composition gradient,[13,14] or an unconventional spin source layer with low crystal or magnetic symmetry.[15-29] Such a lateral asymmetry is typically accompanied by an additional spin current with a $z$-spin component in $\boldsymbol{p}$ ($p_z$), generating out-of-plane damping-like torque $\tau_z\sim\boldsymbol{m}\times(\boldsymbol{m}\times p_z)-$ that can directly reverse the $\boldsymbol{m}$.[30,31,32] Moreover, when $p_z$ is sufficiently strong, the switching current can be significantly reduced since the associated $\tau_z$ does not compete with magnetic anisotropy but directly changes the effective damping in ferromagnets.[31,32]

Recently, studies suggest that asymmetric boundary conditions in a single-layer ferromagnetic film can generate finite $p_z$ by breaking *M* symmetry via reorienting the magnetization to the direction oblique to the film plane.[33,34] This symmetry breaking manifests in spin-dependent transport properties, notably the planar Hall effect (PHE) and anisotropic magnetoresistance (AMR),[33,34,35] which serve as the measure of the *M* symmetry breaking and can induce a spin current flowing and polarized along the magnetization direction, generating a finite $p_z$. However, this approach has not yet been used to switch the perpendicular magnetizations, possibly due to the out-of-plane PHE current only occurring during the thermal fluctuations of SOT-driven dynamics of magnetizations, and thus is not sufficient to generate strong $p_z$ to assist the switching by SHE or REE with an achievable current density.

In this work, we present an effective method for achieving field-free switching of perpendicular magnetization with the assistance of the orbital Hall effect (OHE). This effect produces a transverse orbital current, capable of being converted into a *y*-polarized spin current with greater strength than that generated by conventional SHE and REE. We show that in a Co/PtGd/Mo heterostructure, the Mo layer contributes a strong OHE, amplifying $p_y$ generation through efficient orbital-to-spin conversion. In the presence of a moderate $p_z$ by an enhanced PHE, an efficient and field-free switching assisted by OHE is achieved.

## II. EXPERIMENTS AND RESULTS

The Co(0.8)/Pt(2) and Co(0.8)/PtGd(2)/Mo($a$) films were grown on thermally oxidized Si substrates using magnetron sputtering (thicknesses in parentheses are in nm), where $a$ can be 0, 2, 3, and 4 nm. All stacks featured a 0.5 nm Ta seed layer and a 1.5 nm Pt capping layer to optimize growth and prevent oxidation, respectively. Post-deposition annealing was performed to enhance perpendicular magnetic anisotropy (PMA). The annealing treatment may not only help to improve the quality of the crystalline of the film[36] but also promote the segregation of Co and Gd atoms at the interface layer,[37,38] which will be responsible for the presence of PMA. Devices were fabricated by patterning Hall bars (6 μm current channel width/3 μm voltage arms) followed by Argon ion milling into a 3 μm diameter pillar-shaped device for field-free SOT switching. Details of the sample preparation and measurements are provided in the Methods section.

The PHE is a magnetotransport phenomenon in magnetic materials associated with anisotropic magnetoresistance (AMR), which can be described as $\boldsymbol{J}_{PHE} = \Delta\sigma_{AMR}\boldsymbol{m}(\boldsymbol{m}\cdot\boldsymbol{E})$, where $\boldsymbol{J}_{PHE}$ is the planar Hall current, $\Delta\sigma_{AMR}$ is the conductivity associated with AMR, $\boldsymbol{E}$ is the electric field. The $\boldsymbol{J}_{PHE}$ is then polarized collinear to $\boldsymbol{m}$ and flowing align the direction of $\boldsymbol{m}(\boldsymbol{m}\cdot\boldsymbol{E})$.[33,34,35] The out-of-plane component $\boldsymbol{J}_{PHE}$ ($\boldsymbol{J}_{PHE,\perp}$) can introduce a net spin accumulation of the FM film with the asymmetric boundary condition.[34] We note that for a FM film with $\boldsymbol{m}$ perfectly pointing along the *z* direction, the $\boldsymbol{J}_{PHE,\perp}$ is vanishing. However, $\boldsymbol{m}$ may be pulled away from the *z*-direction by thermal fluctuation or during the SOT-driven dynamics, allowing a nonvanishing $p_y$ in any ferromagnetic films. However, the deterministic switching of perpendicular magnetization associated on PHE has not been demonstrated yet, possibly because PHE in common ferromagnetic metals is insufficient to assist the switching by SHE or REE with an achievable current density. To realize this approach in practice, one needs to find a strategy to efficiently enhance $\Delta\sigma_{AMR}$ and hence the $\boldsymbol{J}_{PHE}$ for a moderate $p_z$, and use a new mechanism beyond SHE or REE to boost $p_y$, as reflected by our magnetic dynamics simulations in supplemental section 1. We propose the PHE can be enhanced by engineering the interfacial exchange coupling between the perpendicular ferromagnet layer and a suitable adjacent layer, and the sufficiently large $p_y$ can be generated by including an additional layer in the SOT device to generate OHE.

We begin by highlighting the enhancement of PHE in the SOT devices illustrated in Fig. (1a), 1(b), and 1(c). A 0.8 nm ferromagnetic Co layer was deposited atop the commonly employed spin source Pt layer with a thickness of 2 nm (denoted as Co(0.8)/Pt(2), as shown in Fig. (1a). The good perpendicular magnetic anisotropy (PMA) of the Co layer is confirmed by the measurement of anomalous Hall effect (AHE) resistance ($R_H$), as shown in Fig. (1d).

Since PHE is proportional to AMR, we measure the longitudinal resistance of this device under a magnetic field rotating within the *x*-*y* and *x*-*z* planes. We find the AMR is relatively small in Co(0.8)/Pt(2), since AMR originates from spin-orbit coupling (SOC), which is small for Co even interfaced with the heavy metal Pt. Considering that the AMR may be enhanced in the presence of a stronger interfacial coupling, we replace the Pt layer with a PtGd alloy layer with the stoichiometry of $Pt_{0.65}Gd_{0.35}$. We find that the perpendicular anisotropy is well maintained with the presence of the PtGd layer, where the $R_H$ is enhanced, as shown in Fig. 1(d). Gd is an element with a significant effective magnetic moment and strong SOC. Although the magnetic order in PtGd alloy is absent,[24] a thin Pt-Gd-Co alloy intermediate layer may appear across the interface between the Co and PtGd layer, which may be ferromagnetic and strongly coupled with the top Co layer,[39] and thus may lead to the different $R_H$ and introduce substantial enhancement of AMR in the Co layer. This is clearly shown in Fig. (1e) and (1f), where the AMR for Co(0.8)/PtGd(2) sample is much stronger than that of Co(0.8)/Pt(2), and is qualitatively consistent with our first-principles calculations (see Fig. S2 in supplemental section 2). We notice that the AMR can be further enhanced in the presence of a Mo layer underneath the PtGd. Recent studies suggest that AMR may be also contributed by the orbital magnetic moments of Bloch electrons.[40,41,42] However, all the AMR measurements for our devices show a 2-fold symmetry, indicating the PHE and AMR can be explained by a conventional phenomenological picture. The influence of orbital magnetic moments may emerge in a magnetic field much stronger than that we applied in this work. In addition, finite interlayer diffusion is common in multilayer systems, which can also bring strong disorder scattering that enhances PHE and hence the $z$-polarized spin current.

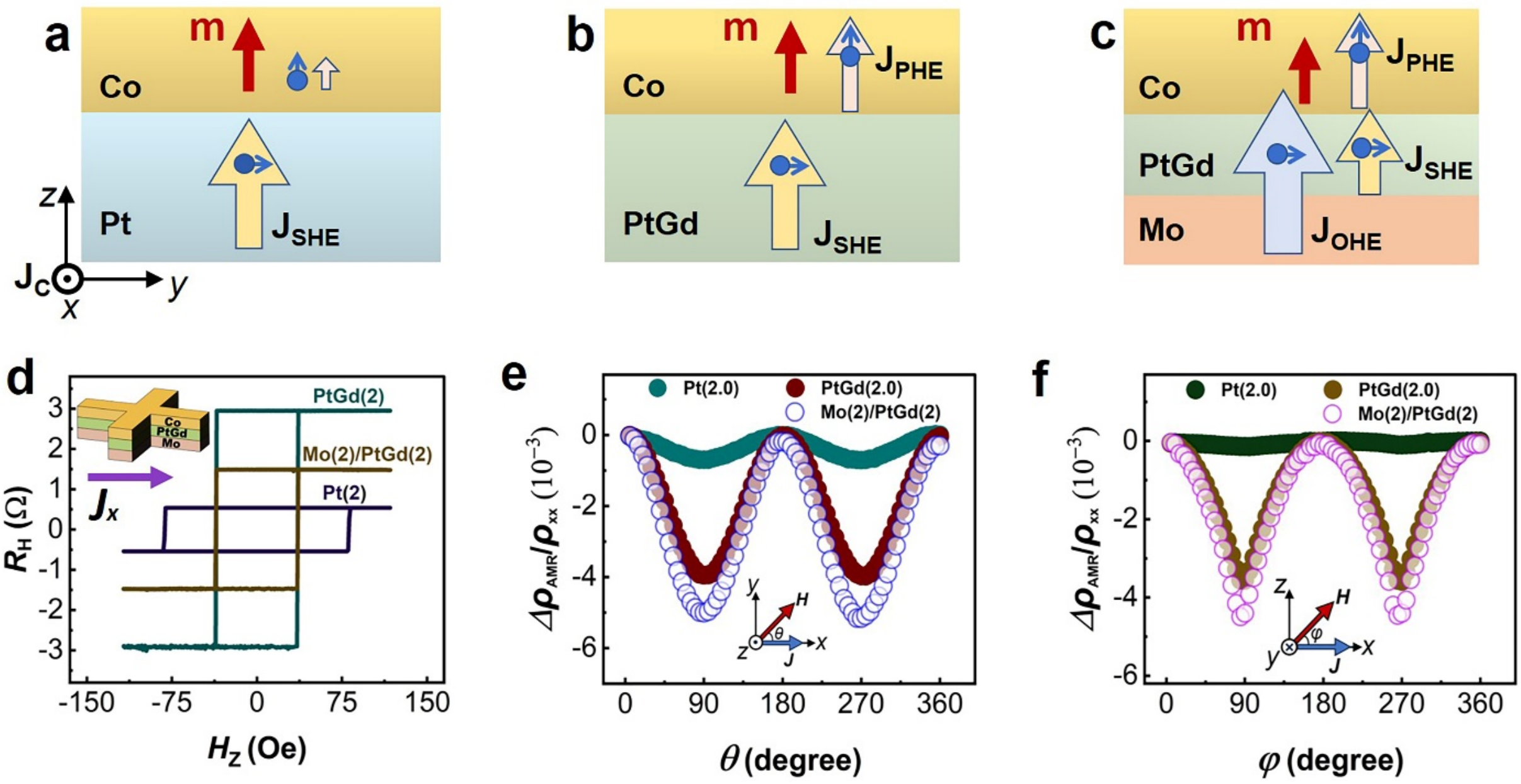

**FIG. 1.** Schematic of Co/PtGd/Mo SOT devices and their basic magneto-transport properties. **(a)** A Co/Pt bilayer SOT device. The Co layer has perpendicular anisotropy. A current $\boldsymbol{J}$ along the $x$ direction generates a negligible out-of-plane $z$-polarized spin current by PHE ($\boldsymbol{J}_{\mathrm{PHE}}$) in Co layer and a moderate out-of-plane $y$-polarized spin current by SHE ($\boldsymbol{J}_{\mathrm{SHE}}$) in the Pt layer. The $\boldsymbol{J}_{\mathrm{SHE}}$ enters the Co layer and exerts a moderate $\tau_y$. (b) A Co/PtGd bilayer SOT device where PtGd supports a moderate $\boldsymbol{J}_{\mathrm{SHE}}$, and the $\boldsymbol{J}_{\mathrm{PHE}}$ in the Co layer is enhanced due to the Co/PtGd interface coupling. The $\boldsymbol{J}_{\mathrm{PHE}}$ can exert an additional $\tau_z$ in Co layer besides $\tau_y$ due to $\boldsymbol{J}_{\mathrm{SHE}}$.

(c) A Co/PtGd/Mo trilayer SOT device. In addition to a not-too-small $\boldsymbol{J}_{\mathrm{PHE}}$ in Co layer and a moderate $\boldsymbol{J}_{\mathrm{SHE}}$ in the PtGd layer, there is a strong orbital current due to OHE $(\boldsymbol{J}_{\mathrm{OHE}})$ in Mo layer flowing along the out-of-plane direction, which will be converted into an additional-*y*-polarized strong spin current when passing through the PtGd layer, and will significantly boost $\tau_y$ in the Co layer. (d) The anomalous Hall resistance $R_H$ as a function of an out-of-plane magnetic field $H_z$ for Co(0.8)/Pt(2), Co(0.8)/PtGd(2), and Co(0.8)/PtGd(2)/Mo(2) Hall bar devices, respectively. The top left inset illustrates the Hall bar device for anomalous Hall measurement. (e-f) Anisotropic magnetoresistivity ratio $(\boldsymbol{\rho}_{\mathbf{AMR}}/\boldsymbol{\rho}_{\mathbf{xx}})$ for Co(0.8)/Pt(2), Co(0.8)PtGd(2), and Co(0.8)/PtGd(2)/Mo(2) samples in the *x*-*y* plane (e) and *x*-*z* plane (f), where the large scanning field of 1.2 T was applied, which is much larger than the anisotropy field $H_k$ about 0.72 T, as shown in Fig. S(3a) in supplemental section 3.

The strongly enhanced AMR (Fig. 1(e) and 1(f)) and PHE (see Fig. S(3b) in supplemental section 3) may result in notable *z*-polarized planar Hall current $(\boldsymbol{J}_{\mathrm{PHE}})$ in the Co layer. To show this, we measure the hysteresis loop of the $R_{\mathrm{H}}$ versus the out-of-plane magnetic field $H_z$ in the presence of a bias dc current $J_x$ along the *x* direction. If there was a finite $p_z$ in $\boldsymbol{J}_{\mathrm{PHE}}$, the associated $\tau_z$ would cause an abrupt shift in the $R_H$-$H_z$ hysteresis loop when $\tau_z$ is sufficiently strong to overcome the intrinsic damping of Co. This approach has been widely used to verify the existence of $p_z$ in SOT devices.[16,17,21,43] We find negligible shift of the loop in the Co(0.8)/Pt(2) sample even for a large $J_x$, as shown in Fig. S(7a) in supplemental section 7. For Co(0.8)/PtGd(2) sample, we find negligible shift for small $J_x$, while sizable positive and negative shifts emerge when $J_x$ is larger (Fig. (2a)), indicating the existence of $p_z$ in Co(0.8)/PtGd(2) sample. We define the shift of the $R_H$-$H_z$ hysteresis loop as $\Delta H_z = H_{center}(J_x^+) - H_{center}(J_x^-)$, where $H_{center}(J_x^\pm) = \frac{[H_r^+(J_x^\pm)+H_r^-(J_x^\pm)]}{2}$ is the centre of the hysteresis loop shifts determined by the difference of positive and negative magnetization-reversal fields, $H_r^+(H_r^-)$ is the rightward (leftward) coercivity field corresponds to a positive (negative) $H_z$ field, and $J_x^\pm$ are the current density of positive and negative currents. Figure (2a) inset curve shows the $\Delta H_z$ as a function of $J_x$, where the $\Delta H_z$ is negligible when $J_x$ is small and increases linearly when $J_x > 11 \times 10^6$ A/cm$^{-2}$ for Co(0.8)/PtGd(2) sample. Similarly, the $R_H$-$H_z$ hysteresis loop shift is observed for Co(0.8)/PtGd(2)/Mo(4) sample, we found sizable positive and negative $R_H$-$H_z$ hysteresis loop shifts emerge when $J_x$ is large (Fig. (2b)). $\Delta H_z$ is negligible when $J_x$ is small and increases linearly when $J_x > 2 \times 10^6$ A/cm$^{-2}$ for Co(0.8)/PtGd(2)/Mo(4) sample (inset of Fig. 2(b)), indicating the enhancement of $p_z$ in Co(0.8)/PtGd(2)/Mo(4) sample. The $R_H$-$H_z$ loops are shifted to the opposite field direction in the absence of $H_x$ field corresponds to the SOT-induced out-of-plane effective field, which supports the field-free SOT switching.[16] This phenomenon cannot occur in conventional SOT devices supporting only $p_y$, where $\Delta H_z$ can only be generated in the presence of an $H_x$ field.[16,21,29] The effective SOT field $\eta_z = \Delta H_z / J_x$ contributed by $p_z$ is then determined with $H_x = 0$ Oe (Fig. (2a), (2b), and (2c)). Subsequently, using a standard harmonic Hall voltage measurement technique, we evaluate the SOT effective field $\eta_y$, as described in Fig. S(8c) in supplemental section 8. For Co(0.8)/PtGd(2) sample, the contribution of $p_z$ from $\boldsymbol{J}_{\mathrm{PHE}}$ results in an out-of-plane SOT field efficiency $\eta_z = 0.17 \times 10^{-6}$ Oe A$^{-1}$ cm$^2$, while the $p_y$ attributed to the spin Hall current ($\boldsymbol{J}_{\mathrm{SHE}}$) yields in-plane SOT field efficiency $\eta_y = 2.01 \times 10^{-6}$ Oe A$^{-1}$ cm$^2$ (Fig. 2(c)). Despite these contributions, both $\eta_y$ and $\eta_z$ may not be strong enough in the device, as evident from the absence of field-free switching of $\boldsymbol{m}$ in Co/PtGd under an achievable $J_x$ (Fig. (3a)). This is in agreement with our previous finding in similar Co/PtGd SOT devices with ultrathin 2 nm PtGd layer.[22]

We then engineer the spin source to enhance $p_y$ and thus the associated $\eta_y$ using the recently discovered OHE, which generated a current of electron orbital angular momentum ($\boldsymbol{J}_{\mathrm{OHE}}$) that flows transverse to a longitudinal current.[44,45,46,47] OHE can exist in materials with weak SOC, such as Mo, Ru, Ti, and Nb.[45,46,47,48] The $\boldsymbol{J}_{\mathrm{OHE}}$ can be converted to a spin current when passing through a layer with strong SOC which can be significantly stronger than $\boldsymbol{J}_{\mathrm{SHE}}$, promising for spin-orbit torque generation.[49] We thus insert a Mo layer in the bottom of the PtGd layer in the SOT device, where an out-of-plane $\boldsymbol{J}_{\mathrm{OHE}}$ is generated by an in-plane charge current $J_x$ in the Mo layer and transformed to a spin current via spin-orbit coupling when flowing through the PtGd layer (Fig. (1c)). We find the $\eta_y$ is enhanced by 2 times and 4 times in the presence of the Mo underlayers with the thickness of 2 nm and 4 nm, respectively, as shown in Fig. (2c). Such significant enhancement of $\eta_y$ clearly proves the efficiency of OHE in the generation of $p_y$. We also measure the $R_H$-$H_z$ hysteresis loop shifts for Co(0.8)/PtGd(2)/Mo($a$) devices and find that the $\eta_z$ is slightly increased in the presence of Mo layer, as shown in Fig. (2c). This is consistence with the enhancement of AMR shown in Fig. (1e) and (1f), proving that the generation of $p_z$ is associated with PHE, as shown in Fig. S(3b) in supplemental section 3.

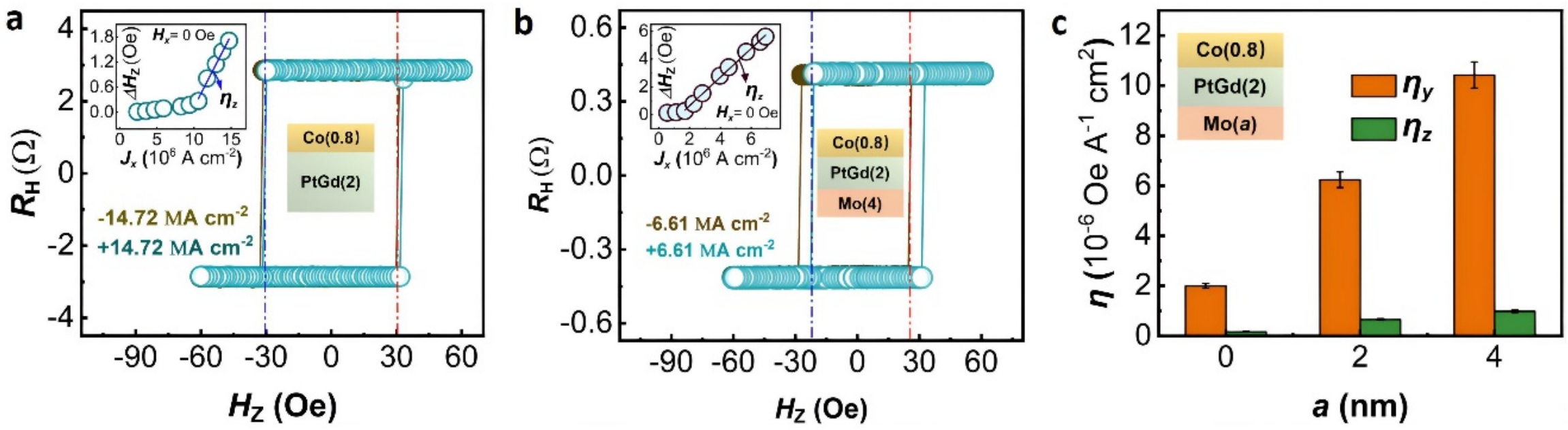

**FIG. 2.** Finite $p_z$ due to PHE and enhanced $p_y$ due to OHE. (a-b) $R_{\mathrm{H}}$ vs $H_Z$ curves for Co(0.8)/PtGd(2) and Co(0.8)/PtGd(2)/Mo(4) Hall bar SOT devices measured by a bias current $J_x$, respectively. For Co(0.8)/PtGd(2) sample (a). For Co(0.8)/PtGd(2)/Mo(4) sample (b). The inset diagram shows the $R_H$ loop shifts $\Delta H_z$ as a function of $J_x$ for Co(0.8)/PtGd(2) and Co(0.8)/PtGd(2)/Mo(4) sample in the absence of $H_x$, respectively. (c) The drive $\tau_y$-associated SOT effective field $\eta_y$ and $\tau_z$- associated SOT effective field $\eta_z$ for Co(0.8)/PtGd(2)/Mo($a$) SOT devices.

We then prepare Co(0.8)/PtGd(2)/Mo($a$) SOT devices with the top Co layer fabricated as pillar-shaped SOT devices to examine the switching behaviour. The thickness of the PtGd layer is fixed to 2 nm to enhance $\boldsymbol{J}_{\mathrm{PHE}}$ and facilitate the conversion of $\boldsymbol{J}_{\mathrm{OHE}}$ into a spin current $\boldsymbol{J}_{\mathrm{SHE}}$. Figure 3 shows the changes in normalized $R_{\mathrm{H}}$ $\left(R_H^{J_x}/R_H^{H_z(S)}\right)$ as a function of an in-plane current $J_x$ applied to these SOT devices. Here, the $R_H^{H_z(S)}$ signifies the saturation anomalous Hall resistance of the pillar-shaped SOT devices, induced by an out-of-plane magnetic field as shown in Fig. S(4a) and S(4b) in supplemental section 4. In contrast to the flat $R_H$-$J_x$ curve observed in the absence of Mo ($a = 0$) (Fig. (3a)), a hysteretic behaviour and a sign change in $R_H$-$J_x$ curve emerges when the Mo layer is introduced ($a > 0$) (Fig. (3b) and 3(c)). In the regime of high $J_C$, the sample undergoes heating and exhibits non-uniform magnetization switching behaviour, as shown in Fig. (3b) and 3(c). This phenomenon is commonly observed in previous studies on deterministic magnetization switching.[50,51,52] The device's temperature rise due to Joule heating is anticipated to be smaller than 37 K at SOT switching current density of 22.52 MA cm$^{-2}$, as illustrated in Fig. S9 in supplemental section 9. This indicates that a temperature effect is not expected to play a dominant role in the field-free switching. Notably, we observed about ~100% field-free magnetization switching in the Co pillar-shaped SOT devices for both $a = 2$ nm and 4 nm (Fig. (3b) and (3c)). This observation reflects a substantial $y$-polarized $\boldsymbol{J}_{\mathrm{OHE}}$ and a finite $z$-polarized $\boldsymbol{J}_{\mathrm{PHE}}$ can result in the field-free switching of perpendicular magnetization.

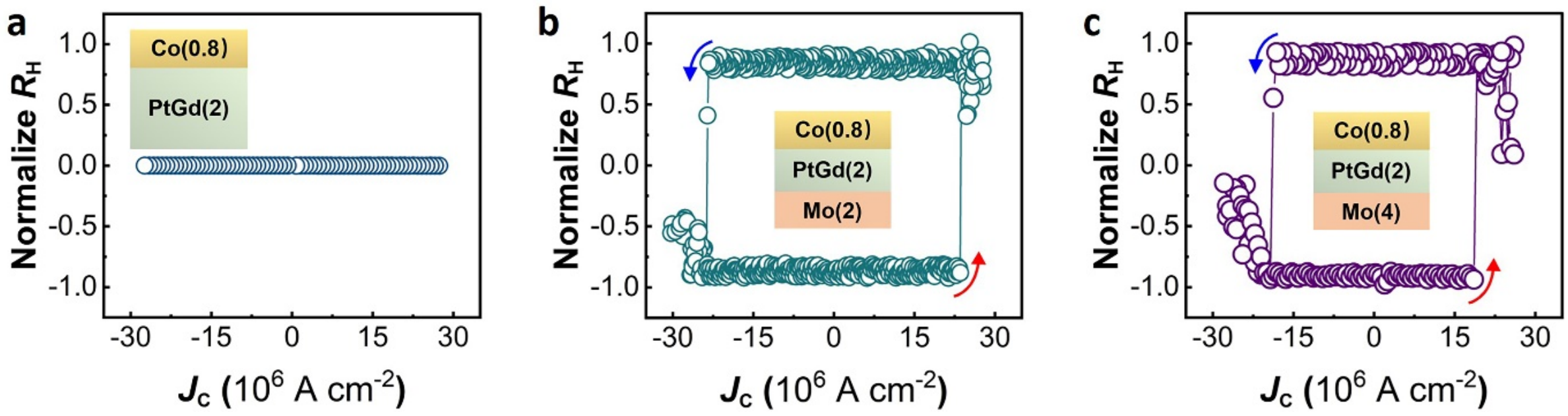

**FIG. 3.** Field-free current-induced magnetization switching in Co(0.8)/PtGd(2)/Mo($a$) pillar-shaped SOT devices. (a-c) The changes of $R_H$ (normalized) as a function of an in-plane current $J_x$ applied of Co(0.8)/PtGd(2)/Mo($a$) SOT devices for $a = 0$ nm (a), $a = 2$ nm (b), and $a = 4$ nm (c). The red and blue arrows show the SOT switching direction as the switching current sweeps from negative to positive and vice versa.

## III. CONCLUSIONS

In conclusion, we have successfully demonstrated the efficient current-induced field-free switching of a perpendicular magnetization in a Co/PtGd/Mo SOT device through the assistance of OHE. In the presence of a PHE-induced moderate $p_z$ and hence the finite out-of-plane damping-like torque in the Co/PtGd bilayer, an additional Mo bottom layer contributes a strong $p_y$ via OHE, generating a significant in-plane damping-like torque. With the combined influence of these SOTs, the perpendicular magnetization of Co is efficiently switched without the need for an external magnetic field. Importantly, this approach does not require a strong symmetry breaking typically induced by engineering the device geometry or utilizing a low symmetric spin source, making it particularly attractive for practical applications. Our findings underscore the promising potential OHE in magnetic switching and provide a straightforward strategy for constructing high-performance and energy-efficient spin-orbitronic devices.

# IV. METHODS

## 1. Sample growth and device fabrication

*Thin Film Preparation:* The Pt(1.5)/Co(0.8)/$Pt_{0.65}Gd_{0.35}$(2)/Mo($a$)/Ta(0.5) and the comparative Pt(1.5)/Co(0.8)/Pt(2)/Ta(0.5) (layers from top to down, thicknesses in nanometers) stack structures were grown on thermally oxidized Si/SiO substrate through magnetron sputtering, where $a$ is 0, 2, 3, and 4 nm. The thin films were fabricated at room temperature under a base pressure of better than $3 \times 10^{-8}$ Torr. The $Pt_{0.65}Gd_{0.35}$ alloy films' stoichiometry was achieved by co-sputtering Pt and Gd-targets using direct current (DC) sputtering powers of 30 and 21 W, respectively. Further details of the stoichiometry $Pt_{0.65}Gd_{0.35}$ alloy preparation and designs can be found in our prior work.[24] All metal layers were deposited using DC sputtering power without breaking the vacuum chamber. A 1.5 nm Pt layer acts as a capping layer to protect against oxidation of the stack structures. The bottom Mo layer served as a source of orbital Hall current, and the 0.5 nm Ta layer was an adhesive layer that promoted the PMA of the Co layer. Cr (10 nm)/Au (60 nm) thick electrical contacts were deposited into a pillar-shaped SOT device after the second lithography. No magnetic field was applied during the thin film deposition process.

Before fabrication into the Hall bar device, the stacks are subjected to pre-annealing at 350°C for 10 min to support PMA. All three samples were annealed without any external magnetic field. The stacked layers are then patterned into Hall bar devices with a current channel width of 6 μm and a voltage arm width of 3 μm. The Hall bar is subsequently patterned into a pillar-shaped SOT device with a diameter of 3 μm using argon ion milling. This process facilitates field-free SOT switching by downscaling the stack and avoiding the contribution of spin currents from the 1.5 nm platinum-capped layer.

## 2. Electrical measurements

By applying a 50 μA current along the Hall bar, the anomalous Hall resistances of the devices as a function of the out-of-plane magnetic field are determined. Keithley 2602B current source and Keithley 2182 nanovoltmeters were used to measure the current-induced deterministic switching loops by injecting pulsed current along the $x$-axis. The harmonic Hall voltage was performed using a Stanford Research SR830 DSP lock-in amplifier with a frequency amplitude of 1327 Hz. All measurements of magnetic and electrical properties were conducted at room temperature.

## SUPPLEMENTARY MATERIAL

See the supplementary material for additional information.

## ACKNOWLEDGEMENTS

This research was supported by the National Key R&D Program of China (Grant Nos. 2022YFA1405100 and 2021YFA1600200), the Beijing Natural Science Foundation for International Scientists Project (Funding No. IS23028), the NSFC (Grant Nos. 12241405, 12104449, 52250418, and 12274411), the strategic priority research program of Chinese Academy of Sciences (Grant Nos. XDPB44000000 and XDB28000000), the Basic Research Program of the Chinese Academy of Sciences Based on Major Scientific Infrastructures (Grant No. JZHKYPT-2021-08), and the CAS Project for Yound Scientists in Basic Research No. YSBR-084.

## AUTHOR DECLARATIONS

### A. Author Contributions

**Wang K:** conceived and supervised this work. **Bekele ZA:** grew the films, fabricated the devices, and carried out the electrical transport measurements. **Jiang YY** and **Shao DF:** performed the theoretical analysis. **Bekele ZA, Shao DF,** and **Wang K:** analyzed the data and wrote the manuscript. All authors discussed the results and commented on the manuscript.

### B. DATA AVAILABILITY

The data that support the findings of this study are available within the article and its supplementary material.

## ETHICS DECLARATIONS

### Competing Interests

The authors have no conflicts to disclose.

## REFERENCES

[1]E. Y. Tsymbal, I. Žutić (Eds.) (2019), "Spintronics Handbook, Second Edition: Spin Transport and Magnetism: Volume Three: Nanoscale Spintronics and Applications (2nd ed.)," CRC Press. https://doi.org/10.1201/9780429441189 (2019).

[2]Y. Cao, A. W. Rushforth, Y. Sheng, H. Zheng, K. Wang, "Tuning a binary ferromagnet into a multistate synapse with spin-orbit-torque-induced plasticity," Adv. Funct. Mater. **29**(25), 1808104 (2019).

[3]L. Liu, O. J. Lee, T. J. Gudmundsen, D. C. Ralph, R. A. Buhrman, "Current-induced switching of perpendicularly magnetized magnetic layers using spin torque from the spin hall effect," Phys. Rev. Lett. **109**(9) 096602 (2012).

[4]L. Liu, C. F. Pai, Y. Li, H. W. Tseng, D. C. Ralph, R. A. Buhrman, "Spin–torque switching with the giant spin Hall effect of tantalum," Science **336**(6081), 555-558 (2012).

[5]I. M. Miron, G. Gaudin, S. Auffret, B. Rodmacq, A. Schuhl, S. Pizzini, J. Vogel, P. Gambardella, "Current-driven spin torque induced by the Rashba effect in a ferromagnetic metal layer," Nat. Mater. **9**(3), 230-234 (2010).

[6]I. M. Miron, K. Garello, G. Gaudin, P.-J. Zermatten, M. V. Costache, S. Auffret, S. Bandiera, B. Rodmacq, A. Schuhl, P. Gambardella, "Perpendicular switching of a single ferromagnetic layer induced by in-plane current injection," Nature **476**(7359), 189 (2011).

[7]S. Fukami, T. Anekawa, C. Zhang, H. Ohno, "A spin-orbit torque switching scheme with collinear magnetic easy axis and current configuration," Nat. Nanotechnol. **11**(7), 621 (2016).

[8]M. Jiang, H. Asahara, S. Ohya, M. Tanaka, "Electric Field Control of Spin–Orbit Torque Magnetization Switching in a Spin–Orbit Ferromagnet Single Layer," Adv. Sci. **10**(24), 2301540 (2023).

[9]Q. Yang, D. Han, S. Zhao, J. Kang, F. Wang, S. -C. Lee, J. Lei, K. -J. Lee, B. -G. Park, H. Yang, "Field-free spin–orbit torque switching in ferromagnetic trilayers at sub-ns timescales," Nat. Commun. **15**(1), 1814 (2024).

[10]S. Fukami, C. Zhang, S. DuttaGupta, A. Kurenkov, H. Ohno, "Magnetization switching by spin-orbit torque in an antiferromagnet-ferromagnet bilayer system," Nat. Mater. **15**(5), 535 (2016).

[11]Y. W. Oh, S. C. Baek, Y. M. Kim, H. Y. Lee, K. D. Lee, C. G. Yang, E. S. Park, K. S. Lee, K. W. Kim, G. Go, J. R. Jeong, B. C. Min, H. W. Lee, K. J. Lee, B. G. Park, "Field-free switching of perpendicular magnetization through spin-orbit torque in antiferromagnet/ferromagnet/oxide structures," Nat. Nanotechnol. **11**(10), 878–884 (2016).

[12]B. Wu, M. Jin, Z. Shao, H. Fan, J. Wen, H. Li, C. Yu, B. Liu, T. Zhou, "Shape anisotropy induced field-free switching and enhancement of dampinglike field in Pt/Co/PtMn heterostructures with a wedged ultrathin antiferromagnetic PtMn layer," Phys. Rev. B **108**(5), 054417 (2023).

[13]Z. Zheng, Y. Zhang, V. Lopez-Dominguez, L. Sánchez-Tejerina, J. Shi, X. Feng, L. Chen, Z. Wang, Z. Zhang, K. Zhang, B. Hong, Y. Xu, Y. Zhang, M. Carpentieri, A. Fert, G. Finocchio, W. Zhao, P. K. Amiri, "Field-free spin-orbit torque-induced switching of perpendicular magnetization in a ferrimagnetic layer with a vertical composition gradient," Nat. Commun. **12**(1), 4555 (2021).

[14]L. Liu, Y. Song, X. Zhao, W. Liu, Z. Zhang, "Full-Scale Field-Free Spin-Orbit Torque Switching in HoCo Structure with a Vertical Composition Gradient," Adv. Sci. **32**(39), 2200328 (2022).

[15]D. MacNeill, G. M. Stiehl, M. H. D. Guimaraes, R. A. Buhrman, J. Park, D. C. Ralph, "Control of spin–orbit torques through crystal symmetry in WTe2/ferromagnet bilayers," Nat. Phys. **13**(3), 300 (2017).

[16]L. Liu, C. Zhou, X. Shu, C. Li, T. Zhao, W. Lin, J. Deng, Q. Xie, S. Chen, J. Zhou, R. Guo, H. Wang, J. Yu, S. Shi, P. Yang, S. Pennycook, A. Manchon, J. Chen, "Symmetry-dependent field-free switching of perpendicular magnetization," Nat. Nanotechnol. **16**(3), 277 (2021).

[17]D. C. Mahendra, D.-F. Shao, V. D.-H. Hou, P. Quarterman, A. Habiboglu, B. Venuti, M. Miura, B. Kirby, A. Vailionis, C. Bi, X. Li, F. Xue, Y.-L. Huang, Y. Deng, S.-J. Lin, W. Tsai, S. Eley, W. Wang, J. A. Borchers, E. Y. Tsymbal, S. X. Wang, "Observation of anti-damping spin-orbit torques generated by in-plane and out-of-plane spin polarizations in $MnPd_3$," Nat. Mater. **22**(5), 591 (2023).

[18]T. Nan, C. X. Quintela, J. Irwin, G. Gurung, D. F. Shao, J. Gibbons, N. Campbell, K. Song, S.-Y. Choi, L. Guo, R. D. Johnson, P. Manuel, R. V. Chopdekar, I. Hallsteinsen, T. Tybell, P. J. Ryan, J.-W. Kim, Y. Choi, P. G. Radaelli, D. C. Ralph, E. Y. Tsymbal, M. S. Rzchowski, C. B. Eom, "Controlling spin current polarization through non-collinear antiferromagnetism," Nat. Commun. **11**(1), 4671 (2020).

[19]X. Chen, S. Shi, G. Shi, X. Fan, C. Song, X. Zhou, H. Bai, L. Liao, Y. Zhou, H. Zhang, A. Li, Y. Chen, X. Han, S. Jiang, Z. Zhu, H. Wu, X. Wang, D. Xue, H. Yang, F. Pan, "Observation of the antiferromagnetic spin Hall effect," Nat. Mater. **20**(6), 800-804 (2021).

[20]A. Bose, N. J. Schreiber, R. Jain, D.-F. Shao, H. P. Nair, J. Sun, X. S. Zhang, D. A. Muller, E. Y. Tsymbal, D. G. Schlom, D. C. Ralph, "Tilted spin current generated by the collinear antiferromagnet ruthenium dioxide," Nat. Electron. **5**(5), 267-274 (2022).

[21]S. Hu, D.-F. Shao, H. Yang, C. Pan, Z. Fu, M. Tang, Y. Yang, W. Fan, S. Zhou, E. Y. Tsymbal, X. Qiu, "Efficient perpendicular magnetization switching by a magnetic spin Hall effect in a noncollinear antiferromagnet," Nat. Commun. **13**(1), 4447 (2022).

[22]C. Cao, S. Chen, R.-C. Xiao, Z. Zhu, G. Yu, Y. Wang, X. Qiu, L. Liu, T. Zhao, D.-F. Shao, Y. Xu, J. Chen, Q. Zhan, "Anomalous anisotropy of spin current in a cubic spin source with noncollinear antiferromagnetism," arXiv:2211.04970 (2022).

[23]Q. Ma, Y. Li, D. B. Gopman, Y. U. P. Kabanov, R. D. Shull, C. L. Chien, "Switching a Perpendicular Ferromagnetic Layer by Competing Spin Currents," Phys. Rev. Lett. **120**(11), 117703 (2018).

[24]Z. A. Bekele, X. Liu, Y. Cao, K. Wang, "High-Efficiency Spin-Orbit Torque Switching Using a Single Heavy-Metal Alloy with Opposite Spin Hall Angles," Adv. Electron. Mater. **7**(1), 2000793 (2020).

[25]Z. A. Bekele, R. Li, Y. Li, Y. Cao, X. Liu, K. Wang, "Tuning the High-Efficiency Field-Free Current-Induced Deterministic Switching via Ultrathin PtMo Layer with Mo Content," Adv. Electron. Mater. **7**(12), 2100528 (2021).

[26]A. M. Humphries, T. Wang, E. R. J. Edwards, S. R. Allen, J. M. Shaw, H. T. Nembach, J. Q. Xiao, T. J. Silva, X. Fan, "Observation of spin-orbit effects with spin rotation symmetry," Nat. Commun. **8**, 911 (2017).

[27]V. P. Amin, J. Zemen, M. D. Stiles, "Interface-generated spin currents," Phys. Rev. Lett. **121**(13), 136805 (2018).

[28]Z. Luo, Q. Zhang, Y. Xu, Y. Yang, X. Zhang, Y. Wu, "Spin-Orbit Torque in a Single Ferromagnetic Layer Induced by Surface Spin Rotation," Phys. Rev. Appl. **11**(6), 064021 (2019).

[29]S. -h. C. Baek, V. P. Amin, Y. -W. Oh, G. Go, S. -J. Lee, G. -H. Lee, K. -J. Kim, M. D. Stiles, B. -G. Park, K. -J. Lee, "Spin currents and spin-orbit torques in ferromagnetic trilayers," Nat. Mater. **17**(6), 509 (2018).

[30]J.C. Slonczewski, "Current-driven excitation of magnetic multilayers," J. Magn. Magn. Mater. **159**(1-2), L1-L7 (1996).

[31]K.-S. Lee, S.-W. Lee, B.-C. Min, K.-J. Lee, "Threshold current for switching of a perpendicular magnetic layer induced by spin Hall effect," Appl. Phys. Lett. **102**(11), 112410 (2013).

[32]D. -K. Lee, K. -J. Lee, "Spin-orbit Torque Switching of Perpendicular Magnetization in Ferromagnetic Trilayers," Sci. Rep. **10**(1), 1772 (2020).

[33]P. He, S. S. -L. Zhang, D. Zhu, S. Shi, O. G. Heinonen, G. Vignale, H. Yang, "Nonlinear Planar Hall Effect," Phys. Rev. Lett. **123**(1), 016801 (2019).

[34]C. Safranski, E. A. Montoya, I. N. Krivorotov, "Spin–orbit torque driven by a planar Hall current," Nat. Nanotechnol. **14**(1), 27-30 (2019).

[35]T. Taniguchi, J. Grollier, M. D. Stiles, "Spin-Transfer Torques Generated by the Anomalous Hall Effect and Anisotropic Magnetoresistance," Phys. Rev. Appl. **3**(4), 044001 (2015).

[36]H. Song, Z. Zhang, X. Liu, Q. Huang, H. Zhou, T. Hou, R. Qi, Z. Zhang, Z. Xin, Z. Wang, "Comparison of thermal stability of Mo/Si multilayers with different crystallinities of Mo layers," Appl. Opt. **62**(10), 2636-2641 (2023)

[37]K. Ueda, A. J. Tan, G. S. D. Beach, "Effect of annealing on magnetic properties in ferrimagnetic GdCo alloy films with bulk perpendicular magnetic anisotropy," AIP Advances **8**(12), 125204 (2018).

[38]S. Vorobiov, Ia. Lytvynenko, T. Hauet, M. Hehn, D. Derecha, A. Chornous, "The effect of annealing on magnetic properties of Co/Gd multilayers," Vacuum **120,** 9-12 (2015).

[39]K. Cai, Z. Zhu, J. M. Lee, R. Mishra, L. Ren, S. D. Pollard, P. He, G. Liang, K. L. Teo, H. Yang, "Ultrafast and energy-efficient spin–orbit torque switching in compensated ferrimagnets," Nat. Electron. **3**, 37-42 (2020).

[40]N. Friedman, "Additional Scattering of Bloch Electrons by Simultaneous Impurity and Lattice Interactions," Phys. Rev. B **7**(4), 1463 (1973)

[41]M. Wu, D. Tu, Y. Nie, S. Miao, W. Gao, Y. Han, X. Zhu, J. Zhou, W. Ning, M. Tian, "Novel π/2-Periodic Planar Hall Effect Due to Orbital Magnetic Moments in $MnBi_2Te_4$," Nano Lett. **22**(1), 73–80 (2022).

[42]R. B. Atencia, A. Agarwal, D. Culcer, "Orbital angular momentum of Bloch electrons: equilibrium formulation, magneto-electric phenomena, and the orbital Hall effect," Arxiv: 2403.07055.

[43]C. F. Pai, M. Mann, A. J. Tan, G. S. D. Beach, "Determination of spin torque efficiencies in heterostructures with perpendicular magnetic anisotropy," Phys. Rev. B **93**(14), 144409 (2016).

[44]B. A. Bernevig, T. L. Hughes, S.-C. Zhang, "Orbitronics: The Intrinsic Orbital Current in p-Doped Silicon," Phys. Rev. Lett. **95**(6), 066601 (2005).

[45]H. Kontani, T. Tanaka, D. Hirashima, K. Yamada, J. Inoue, "Giant Orbital Hall Effect in Transition Metals: Origin of Large Spin and Anomalous Hall Effects," Phys. Rev. Lett. **102**(1), 016601 (2009).

[46]D. Go, D. Jo, C. Kim, H.-W. Lee, "Intrinsic Spin and Orbital Hall Effects from Orbital Texture," Phys. Rev. Lett. **121**(8), 086602 (2018).

[47]Y.-G. Choi, D. Jo, K. -H. Ko, D. Go, K. -H. Kim, H. G. Park, C. Kim, B. -C Min, G. -M. Choi, H. -W. Lee, "Observation of the orbital Hall effect in a light metal Ti," Nature **619**(7968), 52 (2023).

[48]S. Ding, A. Ross, D. Go, L. Baldrati, Z. Ren, F. Freimuth, S. Becker, F. Kammerbauer, J. Yang, G. Jakob, Y. Mokrousov, M. Kläui, "Harnessing Orbital-to-Spin Conversion of Interfacial Orbital Currents for Efficient Spin-Orbit Torques," Phys. Rev. Lett. **125**(17), 177201 (2020).

[49]D. Lee, D. Go, H. -J. Park, W. Jeong, H. -W. Ko, D. Yun, D. Jo, S. Lee, G. Go, J. H. Oh, K. -J. Kim, B. -G. Park, B. -C. Min, H. C. Koo, H. -W. Lee, O. Lee, K. -J. Lee, "Orbital torque in magnetic bilayers," Nat. Commun. **12**(1), 6710 (2021).

[50]R. Yoshimi, K. Yasuda, A. Tsukazaki, k. S. Takahashi, M. Kawaski, Y. Tokura, "Current-driven magnetization switching in ferromagnetic bulk Rashba semiconductor (Ge,Mn)Te," Sci. Adv. **4**(12), eaat9989 (2018).

[51]P. Li, J. Kally, S. S.-L. Zhang, T. Pillsbury, J. Ding, G. Csaba, J. Ding, J. S. Jiang, Y. Liu, R. Sinclair, C. Bi, A. DeMann, G. Rimal, W. Zhang, S. B. Field, J. Tang, W. Wang, O. G. Heinonen, V. Novosad, A. Hoffmann, N. Samarth, M. Wu, "Magnetization switching using topological surface states," Sci. Adv. **5**(8), eaaw3415 (2019).

[52]W. Li, W. Zhu, G. Zhang, H. Wu, S. Zhu, R. Li, E. Zhang, X. Zhang, Y. Deng, J. Zhang, L. Zhao, H. Chang, K. Wang, "Room-Temperature van der Waals Ferromagnet Switching by Spin-Orbit Torques," Adv. Mater. **35**(51), 2303688 (2023).